\documentclass[12pt,a4paper]{iopart}

\expandafter\let\csname equation*\endcsname\relax
\expandafter\let\csname endequation*\endcsname\relax
\usepackage{amsmath}
\usepackage{amssymb}
\usepackage{hyperref}
\usepackage{iopams}

\RequirePackage{color}

\hypersetup{
  colorlinks=true,linkcolor=blue,citecolor=blue,
  filecolor=blue,urlcolor=blue,breaklinks=true
}

\begin{document}

\title{On the Aharonov-Casher scattering in a CPT-odd
  Lorentz-violating background}

\author{F. M. Andrade}
\ead{fmandrade@uepg.br}
\address{
  Departamento de Matem\'{a}tica e Estat\'{i}stica,
  Universidade Estadual de Ponta Grossa,
  84030-900 Ponta Grossa-PR, Brazil
}

\author{E. O. Silva}
\ead{edilbertoo@gmail.com}
\address{
  Departamento de F\'{i}sica,
  Universidade Federal do Maranh\~{a}o,
  Campus Universit\'{a}rio do Bacanga,
  65085-580 S\~{a}o Lu\'{i}s-MA, Brazil
}
\author{T. Prud\^{e}ncio}
\ead{thprudencio@gmail.com}
\address{
  Instituto de F\'{i}sica,
  Universidade de Bras\'{i}lia,
  Caixa Postal 04455,
  70919-970, Bras\'ilia-DF, Brazil.
}

\author{C. Filgueiras}
\ead{cleversonfilgueiras@yahoo.com.br}
\address{
  Departamento de F\'{i}sica,
  Universidade Federal de Campina Grande,
  Caixa Postal 10071,
  58109-970 Campina Grande-PB, Brazil
}

\date{\today}

\begin{abstract}
The effects of a Lorentz symmetry violating background vector on
the Aharonov-Casher scattering in the nonrelativistic limit
is considered.
By using the self-adjoint extension method we found that there
is an additional scattering for any value of the self-adjoint
extension parameter and non-zero energy bound states for
negative values of this parameter.
Expressions for the energy bound states, phase-shift and the
scattering matrix are explicitly determined in terms of the
self-adjoint extension parameter.
The expression obtained for the scattering amplitude reveals
that the helicity is not conserved in this scenario.
\end{abstract}

\pacs{11.30.Cp, 03.65.Ge, 03.65.Db, 11.55.-m}
\submitto{\jpg}
\maketitle

\section{Introduction}
\label{sec:introduction}

The standard model extension (SME) proposed by Colladay and
Kosteleck\'{y}
\cite{PRD.1997.55.6760,PRD.1998.58.116002,PRD.2004.69.105009}
(cf. also Refs.
\cite{PRD.1999.59.116008,PRL.1989.63.224,PRL.1991.66.1811}) has
been an usual framework for investigating signals of Lorentz
violation in physical systems and has inspired a great deal of
investigations in this theme in recent years.
The interest in this issue appears in the different contexts,
such as field theory
 \cite{PRL.1999.82.3572,PRL.1999.83.2518,PRD.1999.60.127901,
PRD.2001.63.105015,PRD.2001.64.046013,JPA.2003.36.4937,
PRD.2006.73.65015,PRD.2009.79.123503,PD.2010.239.942,
PRD.2012.86.065011,PRD.2008.78.125013,PRD.2011.84.076006,
EPL.2011.96.61001,PRD.2012.85.085023,PRD.2012.85.105001,
EPL.2012.99.21003,PRD.2011.84.045008,arXiv:1304.2090}
aspects on the gauge sector of the SME
\cite{NPB.2003.657.214,NPB.2001.607.247,PRD.1995.51.5961,
PRD.1998.59.25002,PLB.1998.435.449,PRD.2003.67.125011,
PRD.2009.80.125040},
quantum electrodynamics
\cite{EPJC.2008.56.571,PRD.2010.81.105015,PRD.2011.83.045018,
EPJC.2012.72.2070,PRD.2012.86.045003,JPG.2012.39.125001,JPG.2012.39.35002},
and astrophysics
\cite{PRD.2002.66.081302,AR.2009.59.245,PRD.2011.83.127702}.
These many contributions have elucidated effects induced by
Lorentz violation and served to set up stringent upper bounds on
the Lorentz-violating (LV) coefficients \cite{RMP.2011.83.11}.
The physical properties of the physical systems can be accessed by
including in all sectors of the minimal standard model LV terms.
In the fermion sector, for example, this violation is
implemented by introducing two CPT-odd terms,
$V_{\mu}\bar{\psi}\gamma ^{\mu }\psi $, $W_{\mu}\bar{\psi}
\gamma_{5}\gamma ^{\mu }\psi $, where $V_{\mu}$, $W_{\mu}$
are the LV backgrounds.
The LV terms are generated as vacuum expectation values of
tensors defined in a high energy scale.
The SME has also been used as a
framework to propose Lorentz violation
\cite{PRL.2001.87.251304,PRD.2002.66.056005} and CPT
\cite{PRL.1997.79.1432,PRD.1998.57.3932,PRL.2002.88.090801,
PRL.2000.84.1381,PRL.2000.84.1098,PRL.1999.82.2254} probing
experiments, which have amounted to the imposition of stringent
bounds on the LV coefficients.
By carefully analyzing the sectors of the SME some authors have
specialized in introducing news nonminimal couplings between
fermionic and gauge fields in the context of the Dirac equation.
In Ref. \cite{EPJC.2005.41.421}, for example, a LV
and CPT-odd nonminimal coupling between fermions and the gauge
field was proposed in the form
\begin{equation}
  D_{\mu}=\partial_{\mu}+ieA_{\mu}+i\frac{g}{2}
  \epsilon_{\mu\lambda\alpha\beta}\upsilon^{\lambda}F^{\alpha\beta},
  \label{eq:cov}
\end{equation}
in the context of the Dirac equation,
\begin{equation}
  (i\gamma^{\mu}D_{\mu}-M)\Psi=0,
  \label{eq:dirac1}
\end{equation}
were $\Psi $ is the fermion spinor,
$\upsilon^{\mu}=(\upsilon_{0},\boldsymbol{\upsilon})$ is the
Carroll-Field-Jackiw four-vector, $g$ is a constant that
measures the nonminimal coupling magnitude, and $F^{\mu\nu}$
is the electromagnetic field tensor, with
\begin{equation}
  F^{0i}=-F^{i0}=E^{i},~~~F^{ij}=-F^{ji}=\epsilon ^{ijk}\,B_{k}.
  \label{eq:fields}
\end{equation}
This suggests that the LV background, intervening in spacetime,
may correct or generate some new properties to the particles.
The analysis of the nonrelativistic limit of
Eq. \eqref{eq:dirac1} reveals that the nonminimal coupling of the
background with the electromagnetic fields generates a magnetic
dipole moment $g\boldsymbol{\upsilon}$ even for non-charged particles
\cite{EPJC.2005.41.421}, yielding an Aharonov-Casher (AC) phase
\cite{PRL.1984.53.319} for its wave function.
The nonminimal coupling in Eq. \eqref{eq:cov} has been applied to
several physical systems in relativistic and nonrelativistic quantum
mechanics
\cite{PRD.2012.86.045001,ADP.2011.523.910,JMP.2011.52.063505,
PRD.2011.83.125025,PLB.2006.639.675,EPJC.2009.62.425,
JPG.2012.39.055004,JPG.2012.39.105004,AP.2013.333.272,
JPG.2012.39.085001,JPG.2013.40.065002}.
Recently, a new CPT-even and LV dimension-five nonminimal
coupling between fermionic and gauge fields, involving the
CPT-even and Lorentz-violating gauge tensor of the SME was
proposed in Ref. \cite{PRD.2013.87.047701}
(cf. also Ref. \cite{PRD.2012.86.125033}).
This new nonminimal coupling was identified by
\begin{equation}
  D_{\mu}=\partial_{\mu}+ieA_{\mu}+\frac{\lambda}{2}
  \left(K_{F}\right)_{\mu\nu\alpha\beta }
  \gamma^{\nu}F^{\alpha\beta},
  \label{eq:covader}
\end{equation}
with $\left(K_{F}\right)_{\mu\nu\alpha\beta}$ being the
tensor ruling  the Lorentz violation in the CPT-even
electrodynamics of the SME.
This nonminimal coupling modifies the Dirac equation, whose
nonrelativistic regime is governed by a Hamiltonian which
induces new effects, such as an electric-Zeeman-like spectrum
splitting and an anomalous-like contribution to the electron
magnetic moment, among others.

In this paper, we specialize to the nonminimal coupling in
Eq. \eqref{eq:cov}.
The aim is to study the effects of this LV background in the scattering
process of a spin-1/2 neutral particle with magnetic dipole moment
$g\boldsymbol{\upsilon}$ in the presence of a electric field of an
infinitely long, infinitesimally thin line of charge, 
in the nonrelativistic limit.

The work is outlined in the following way: In Section
\ref{sec:motion} we derive the Schr\"{o}dinger-Pauli equation in
order to study the physical implications of the LV background on
the spin-1/2 AC scattering problem.
The Section \ref{sec:selfae} is devoted to the study of the LV
Hamiltonian via the self-adjoint extension technique and are
presented some important properties of the LV wave function.
In Section \ref{sec:scattbound} are addressed the scattering and
bound-state problems within the framework of the LV
Schr\"{o}dinger-Pauli equation.
Expressions for the energy bound-states, phase-shift and the
scattering matrix are computed and all them are explicitly
described in terms of the physical condition of the problem and,
as it was expected, the self-adjoint extension parameter is also
expressed in terms of the physical parameters.
At the end, we make a detailed analysis of the helicity
conservation's problem in the present framework.
In Section \ref{sec:conclusion} we give our conclusions and
remarks.

\section{The equation of motion}
\label{sec:motion}

In this section, we derive the equation of motion that governs
the dynamics of a spin-1/2 neutral particle in a radial electric
field and a LV background.
We start with the (2+1)-dimensional Dirac equation, which
follows from the decoupling of (3+1)-dimensional Dirac equation
for the specialized case where $\partial _{3}=0$, into two
uncoupled two-component equations, such as implemented in
Refs. \cite{PRD.1978.18.2932,NPB.1988.307.909,PRL.1989.62.1071}.
Since we are interested only in the effects of the LV
background, we can consider only the sector generating the AC
effect in Eq. \eqref{eq:cov}.
In this case, the planar Dirac equation ($\hbar=c=1$) is
\begin{equation}
  \left(
    \beta \mathbf{\gamma }\cdot \boldsymbol{\Pi}+\beta M
  \right) \Psi =
  \bar{E} \Psi,
\label{eq:defdirac}
\end{equation}
where $\Psi$ is a two-component spinor,
$\boldsymbol{\Pi}=\mathbf{p}-
gs\left(\boldsymbol{\upsilon}\times \mathbf{E}\right)$
is the generalized momentum, and $s$ is twice the spin value,
with $s=+1$ for spin ``up'' and $s=-1$ for spin ``down''.
The $\gamma$-matrices in $(2+1)$ dimensions are given in terms of the
Pauli matrices
\begin{equation}
  \beta = \gamma _{0}=\sigma _{3},\qquad
  \gamma_{1}=i\sigma _{2},\qquad
\gamma_{2}=-is\sigma _{1}.
\end{equation}
The field configuration (in cylindrical coordinates) is chosen
to be
\begin{equation}
  \mathbf{E}=2\lambda
  \frac{\mathbf{\hat{r}}}{r},\qquad
  \boldsymbol{\nabla }\cdot\mathbf{E}=
  2\lambda\frac{\delta(r)}{r},\qquad
  \upsilon^{\mu}=(0,0,0,\upsilon),
  \label{eq:fieldac}
\end{equation}
where $\mathbf{E}$ is the electric field generated by an infinite
charge filament and $\lambda$ is the linear charge density along the
$z$-axis.
The second order equation implied by Eq. \eqref{eq:defdirac} is
obtained by applying the matrix operator
$\left[M+\beta\bar{E}-
\mathbf{\gamma}\cdot\boldsymbol{\Pi}\right]\beta$.
After this application, one finds
\begin{align}
  \left(\bar{E}^{2}-M^{2}\right)\Psi
  = {} & -
  \left(\mathbf{\gamma}\cdot\boldsymbol{\Pi}\right)
  \left(\mathbf{\gamma}\cdot\boldsymbol{\Pi}\right)
  \Psi \nonumber \\
  = {} &
  \left\{
    \boldsymbol{\Pi}^{2}+g\mathbf{\sigma }\cdot
    \left[
      \boldsymbol{\nabla}\times
      \left(\boldsymbol{\upsilon}\times \mathbf{E}\right)
    \right]
  \right\} \Psi.
\label{eq:hrel}
\end{align}
By accessing the nonrelativistic limit,
$\bar{E}=M+E$, $M\gg E$, we obtain the
Schr\"{o}dinger-Pauli equation
\begin{equation}
  \hat{H}\Psi =E\Psi,
  \label{eq:dedfm}
\end{equation}
with
\begin{equation}
  \hat{H}=\frac{1}{2M}
  \left[
    \mathbf{p}-gs\left(\boldsymbol{\upsilon}\times \mathbf{E}\right)
  \right]^{2}+
  \frac{1}{2M}g\mathbf{\sigma}\cdot
  \left[
    \boldsymbol{\nabla }\times(\boldsymbol{\upsilon}\times \mathbf{E})
  \right],
  \label{eq:hdef}
\end{equation}
the Hamiltonian operator.
Using \eqref{eq:fieldac}, the Hamiltonian \eqref{eq:hdef}
becomes
\begin{equation}
  \hat{H}=
  \frac{1}{2M}
  \left[
  \hat{H}_{0} +
  \alpha \sigma_{z} \frac{\delta(r)}{r}
  \right],
  \label{eq:hnrf}
\end{equation}
with
\begin{equation}
    \label{eq:hzero}
  \hat{H}_{0}=
  \left(
    \frac{1}{i}\boldsymbol{\nabla}
    -\alpha s\frac{\hat{\boldsymbol{\varphi}}}{r}
  \right)^{2},
\end{equation}
and
\begin{equation}
\alpha = 2g \upsilon\lambda,
\label{eq:deltac}
\end{equation}
is the coupling constant of the $\delta(r)/r$ potential.

The Hamiltonian in Eq. \eqref{eq:hnrf} governs the quantum dynamics of a
spin-1/2 neutral particle with a radial electric field, i.e., a spin-1/2
AC problem, with $g\boldsymbol{\upsilon}$ playing the role of a
nontrivial magnetic dipole moment, in contrast with the usual AC problem
where the magnetic dipole moment is
$\boldsymbol{\mu}=\mu\sigma_{z} \hat{\mathbf{z}}$
\cite{EPJC.2013.73.2402}.
Also, in Eq. \eqref{eq:hnrf} we observe the presence of a $\delta$
function which is singular at the origin.
This makes the problem more complicated to be solved.
Such kind of point interaction potential can then be addressed
by the self-adjoint extension approach
\cite{JMP.1985.26.2520,Book.2004.Albeverio}, which will be used
for studying the scattering and bound state scenarios.

\section{Self-adjoint extension analysis}
\label{sec:selfae}

An operator $\mathcal{O}$, with domain
$\mathcal{D}(\mathcal{O})$, is said to be self-adjoint if and
only if
$\mathcal{D}(\mathcal{O}^{\dagger})=\mathcal{D}(\mathcal{O})$
and $\mathcal{O}^{\dagger}=\mathcal{O}$.
In order to determine all self-adjoint extensions of
\eqref{eq:hzero}, making use of the underlying rotational
symmetry expressed by the fact that $[\hat{H},\hat{J}_{z}]=0$,
where $\hat{J}_{z}=-i\partial/\partial_{\varphi}+\sigma_{z}/2$ is
the total angular momentum operator in the $z$-direction,
we decompose the Hilbert space
$\mathfrak{H}=L^{2}(\mathbb{R}^{2})$
with respect to the total angular momentum
$\mathfrak{H}=\mathfrak{H}_{r}\otimes\mathfrak{H}_{\varphi}$,
where $\mathfrak{H}_{r}=L^{2}(\mathbb{R}^{+},rdr)$ and
$\mathfrak{H}_{\varphi}=L^{2}(\mathcal{S}^{1},d\varphi)$,
with $\mathcal{S}^{1}$ denoting the unit sphere in
$\mathbb{R}^{2}$.
So, it is possible to express the eigenfunctions of the two
dimensional Hamiltonian in terms of the eigenfunctions of
$\hat{J}_{z}$
\begin{equation}
  \Psi(r,\varphi)=
  \left(
    \begin{array}{c}
      \psi_{m}(r) e^{i(m_{j}-1/2)\varphi } \\
      \chi_{m}(r) e^{i(m_{j}+1/2)\varphi }
    \end{array}
  \right) ,
\label{eq:wavef}
\end{equation}
with $m_{j}=m+1/2=\pm 1/2,\pm 3/2,\ldots $, and $m\in
\mathbb{Z}$.
By inserting Eq. \eqref{eq:wavef} into Eq. \eqref{eq:dedfm} the
Schr\"{o}dinger-Pauli equation for $\psi_{m}(r)$ is found to be
($k^{2}=2ME$)
\begin{equation}
  H\psi_{m}(r)=k^{2}\psi_{m}(r),
  \label{eq:eigen}
\end{equation}
where
\begin{equation}
  H=H_{0}+\alpha \frac{\delta(r)}{r},
  \label{eq:hfull}
\end{equation}
and
\begin{equation}
  H_{0}=
  -\frac{d^{2}}{dr^{2}}-\frac{1}{r}\frac{d}{dr}
  +\frac{(m-\alpha s)^{2}}{r^{2}}.
\end{equation}

The self-adjoint extension approach consists, essentially, in
extending the domain $\mathcal{D}(H_{0})$ to match
$\mathcal{D}(H_{0}^{\dagger})$ and therefore turning $H_{0}$
into a self-adjoint operator.
To do so, we must find the deficiency subspaces,
$N_{\pm}$, with dimensions $n_{\pm}$, which are called
deficiency indices of $H_{0}$ \cite{Book.1975.Reed.II}.
A necessary and sufficient condition for $H_{0}$ being
essentially self-adjoint is that $n_{+}=n_{-}=0$.
On the other hand, if $n_{+}=n_{-}\geq 1$, then $H_{0}$ has an
infinite number of self-adjoint extensions parametrized by
unitary operators  $U:N_{+}\to N_{-}$.
In order to find the deficiency subspaces of $H_{0}$ in
$\mathfrak{H}_{r}$, we must solve the eigenvalue equation
\begin{equation}
  H_{0}^{\dagger}\psi_{\pm}
  =\pm i k_{0}^{2} \psi_{\pm},
  \label{eq:eigendefs}
\end{equation}
where $k_{0}^{2}\in\mathbb{R}$ was introduced for dimensional
reasons.
Since $H_{0}^{\dagger}=H_{0}$, the solutions of
Eq. \eqref{eq:eigendefs} which vanishes at the infinite are the
Hankel functions (up to a constant)
\begin{equation}
\psi_{\pm}=H_{|m-\alpha s|}^{(1)}(\sqrt{\mp i} k_{0} r),
\label{eq:psidefs}
\end{equation}
with $\rm{Im} \sqrt{\pm i}>0$.
The dimension of such deficiency subspace is thus
$(n_{+},n_{-})=(1,1)$.
According to the von Neumann-Krein theory, all self-adjoint
extensions $H_{\theta,0}$ of $H_{0}$ are given by the
one-parameter family
\begin{equation}
\mathcal{D}(H_{\theta,0})=\mathcal{D}(H_{0}^{\dagger})=
\mathcal{D}(H_{0})\oplus N_{+}\oplus N_{-}.
\end{equation}
Thus, $\mathcal{D}(H_{\theta,0})$ in $\mathfrak{H}_{r}$ is given
by the set of functions \cite{Book.1975.Reed.II}
\begin{equation}
  \label{eq:domain}
  \psi_{\theta}(r) = \psi_{m}(r) +
  c\left[ H_{|m-\alpha s|}^{(1)}(\sqrt{-i} k_{0}r)+
    e^{i\theta}H_{|m-\alpha s|}^{(1)}(\sqrt{i}k_{0}r) \right] ,
\end{equation}
where $\psi_{m}(r)$, with $\psi_{m}(0)=\dot{\psi}_{m}(0)=0$
($\dot{\psi}\equiv d\psi/dr$), is the regular wave function,
$c\in \mathbb{C}$ and the number $\theta \in [0,2\pi)$
represents a choice for the boundary condition.
Using the unitary operator
$U: L^{2}(\mathbb{R}^{+},rdr) \to L^{2}(\mathbb{R}^{+}, dr)$,
given by $(U \xi)(r)=r^{1/2}\xi(r)$, the operator $H_{0}$
becomes
\begin{equation}
  \tilde{H}_{0}=
  U H_{0} U^{-1}=
  -\frac{d^{2}}{dr^{2}}
  +\frac{(m-\alpha s)^{2}-1/4}{r^{2}}.
\end{equation}
By standard results the radial operator $\tilde{H}_{0}$,
is essentially self-adjoint for $|m-\alpha s| \geq 1$, while
for $|m-\alpha s|< 1$ it admits an one-parameter family of
self-adjoint extensions \cite{Book.1975.Reed.II}.
This statement can be understood based in
Eq. \eqref{eq:psidefs}, because for $|m-\alpha s| \geq 1$ the
right hand side is not in $\mathfrak{H}_{r}$ at $0$, while it is
in  $\mathfrak{H}_{r}$ for $|m-\alpha s|<1$.

All the self-adjoint extensions $H_{0,\lambda_{m}}$ of
$\tilde{H}_{0}$ are parametrized by the boundary condition at
the origin \cite{JMP.1985.26.2520,Book.2004.Albeverio}
\begin{equation}
\psi^{(0)}=\lambda_{m}\psi^{(1)},  \label{eq:bc}
\end{equation}
with
\begin{align*}
  \psi^{(0)}={} & \lim_{r\rightarrow 0^{+}}r^{|m-\alpha s|}\psi_{m}(r), \\
  \psi^{(1)}={} & \lim_{r\rightarrow 0^{+}}\frac{1}{r^{|m-\alpha s|}}
  \left[\psi_{m}(r)-\psi^{(0)}\frac{1}{r^{|m-\alpha s|}}\right],
\end{align*}
where $\lambda_{m}$ is the self-adjoint extension parameter.
In \cite {Book.2004.Albeverio} is shown that there is a
relation between the self-adjoint extension parameter
$\lambda_{m}$ and the number $\theta$ in Eq. \eqref{eq:domain}.
The number $\theta$ is associated with the mapping of
deficiency subspaces and extend the domain of operator to make
it self-adjoint.
The self-adjoint extension parameter $\lambda_{m}$ have a
physical interpretation: it represents the scattering length
\cite{Book.2011.Sakurai} of $H_{0,\lambda_{m}}$
\cite{Book.2004.Albeverio}.
For $\lambda_{m}=0$ we have the free Hamiltonian (without the
$\delta$ function) with regular wave functions at origin and
for $\lambda_{m}\neq 0$ the boundary condition in Eq. \eqref{eq:bc}
allows a $r^{-|m-\alpha s|}$ singularity in the wave functions at
origin.

\section{Scattering and bound state analysis}
\label{sec:scattbound}

The general solution for Eq. \eqref{eq:eigen} in the $r\neq 0$
region can be written as
\begin{equation}
\label{eq:sol1}
\psi_{m}(r)=a_{m}J_{|m-\alpha s|}(kr)+b_{m}Y_{|m-\alpha s|}(kr),
\end{equation}
with $a_{m}$ and $b_{m}$ being constants and $J_{\nu}(z)$ and
$Y_{\nu}(z)$ are the Bessel functions of first and second kind,
respectively.
Upon replacing $\psi_{m}(r)$ in the boundary condition
\eqref{eq:bc}, one obtain
\begin{equation}
  \lambda_{m} a_{m} \mathcal{A} k^{|m-\alpha s|}=
  b_{m}\bigg[\mathcal{B} k^{-|m-\alpha s|}
  -\lambda_{m}\Big(\mathcal{C}  k^{|m-\alpha s|}
  + \mathcal{B} \mathcal{D} k^{-|m-\alpha s|}
  \lim_{r\rightarrow 0^{+}}r^{2-2|m-\alpha s|}\Big)\bigg],
  \label{eq:bcf}
\end{equation}
with
\begin{align}
\mathcal{A} & = \frac{1}{2^{|m-\alpha s|}\Gamma(1+|m-\alpha
  s|)}, &
\mathcal{B} & =-\frac{2^{|m-\alpha s|}\Gamma(|m-\alpha s|)}{\pi},
\nonumber \\
\mathcal{C} &=-\frac{\cos (\pi |m-\alpha s|)
 \Gamma(-|m-\alpha s|)}{\pi 2^{|m-\alpha s|}}, &
\mathcal{D} &= \frac{k^{2}}{4(1-|m-\alpha s|)}.
\end{align}
In Eq. \eqref{eq:bcf}, $\lim_{r\rightarrow 0^{+}}r^{2-2|m-\alpha s|}$
is divergent if $|m-\alpha s|\geq 1$, hence $b_{m}$ must be zero.
On the other hand, $\lim_{r\rightarrow 0^{+}}r^{2-2|m-\alpha s|}$
is finite for $|m-\alpha s|<1$.
This means that there arises the contribution of the irregular
solution $Y_{|m-\alpha s|}(kr)$.
Here, the presence of an irregular solution contributing to the
wave function stems from the fact the Hamiltonian $H_{0}$ is not a
self-adjoint operator when $|m-\alpha s|<1$ (cf. Section
\ref{sec:selfae}), hence such irregular solution must be
associated with a self-adjoint extension of the operator $H_{0}$
\cite{JPA.1995.28.2359,PRA.1992.46.6052}.
Thus, for $|m-\alpha s|<1$, we have
\begin{equation}
  \lambda_{m} a_{m}\mathcal{A} k^{|m-\alpha s|}=
  b_{m}(\mathcal{B} k^{-|m-\alpha s|}-
  \lambda_{m}\mathcal{C} k^{|m-\alpha s|}),
\end{equation}
and by substituting the values of $\mathcal{A}$, $\mathcal{B}$
and $\mathcal{C}$ into above expression we find
\begin{equation}
  b_{m}=-\mu_{m}^{\lambda_{m}}(k,\alpha) a_{m},
\end{equation}
where
\begin{equation}
  \mu_{m}^{\lambda_{m}}(k,\alpha)=
  \frac
  {\lambda_{m} k^{2|m-\alpha s|}
    \Gamma(1-|m-\alpha s|)\sin(\pi |m-\alpha s|)}
  {\lambda_{m} k^{2|m-\alpha s|}
  \Gamma{(1-|m-\alpha s|)}\cos(\pi|m-\alpha s|)+
  4^{|m-\alpha s|}\Gamma(1+|m-\alpha s|)}.
  \label{eq:mul}
\end{equation}
Since a $\delta$ function is a very short range potential, it
follows that the asymptotic behavior of $\psi_{m}(r)$ for
$r\rightarrow \infty$ is given by \cite{JPA.2010.43.354011}
\begin{equation}
  \psi_{m}(r)\sim \sqrt{\frac{2}{\pi kr}}
  \cos \left[ kr-\frac{|m|\pi}{2}-
  \frac{\pi}{4}+
  \delta_{m}^{{\lambda_{m}}}(k,\alpha)\right] ,
\label{eq:f1asim}
\end{equation}
where $\delta_{m}^{{\lambda_{m}}}(k,\alpha)$  is a
scattering phase shift.
The phase shift is a measure of the argument difference to the
asymptotic behavior of the solution $J_{|m|}(kr)$ of the radial
free equation which is regular at the origin.
By using the asymptotic behavior of the Bessel functions
\cite{Book.1972.Abramowitz} into Eq. \eqref{eq:sol1} one obtain
\begin{align}
  \label{eq:scattsol}
  \psi_{m}(r)
  \sim {}
  &
  a_{m}
  \sqrt{\frac{2}{\pi kr}}
  \left\{
    \cos\left[kr-\frac{\pi |m-\alpha s|}{2}-\frac{\pi}{4}\right]
  \right.
  \nonumber \\
  &
  \left.
    -\mu_{m}^{{\lambda_{m}}}(k,\alpha)
    \sin\left[ kr-\frac{\pi |m-\alpha s|}{2}-\frac{\pi}{4}\right]
  \right\} .
\end{align}
By comparing the above expression with Eq. \eqref{eq:f1asim}, we
have
\begin{equation}
  \cos
  \left[
    kr-\frac{\pi |m-\alpha s|}{2}-\frac{\pi}{4}+
    \theta_{m}^{\lambda_{m}}(k,\alpha)
  \right]=
  \cos
  \left[
    kr-\frac{\pi|m|}{2}-\frac{\pi}{4}+
    \delta_{m}^{{\lambda_{m}}}(k,\alpha)
  \right].
\end{equation}
with $\theta_{m}^{\lambda_{m}}(k,\alpha)$ given by
\begin{equation}
  \cos \left[\theta_{m}^{\lambda_{m}}(k,\alpha)\right] = a_{m}, \qquad
      \sin\left[\theta_{m}^{\lambda_{m}}(k,\alpha)\right] = a_{m}
    \mu_{m}^{{\lambda_{m}}}(k,\alpha).
    \label{eq:sincostheta}
\end{equation}
Thus, the asymptotic behavior in Eq. \eqref{eq:f1asim} is satisfied if
\begin{equation}
a_{m}=\frac{1}{\sqrt{1+\left[\mu_{m}^{{\lambda_{m}}}(k,\alpha)\right]^{2}}}.
\end{equation}
Now, comparing the arguments of the cosines above, the sought phase
shift is obtained
\begin{equation}
\delta_{m}^{{\lambda_{m}}}(k,\alpha)=
\Delta_{m}(\alpha)+\theta_{m}^{\lambda_{m}}(k,\alpha),
\label{eq:phaseshift}
\end{equation}
with
\begin{equation}
\Delta_{m}(\alpha)=\frac{\pi}{2}(|m|-|m-\alpha s|),
\end{equation}
the phase shift of the AC scattering and
\begin{equation}
  \theta_{m}^{\lambda_{m}}(k,\alpha)=\arctan {[\mu_{m}^{\lambda_{m}}(k,\alpha)]}.
\end{equation}
Therefore, by using Eq. \eqref{eq:sincostheta}, the scattering operator
$S_{m}^{\lambda_{m}}(k,\alpha)$ ($S$-matrix) for the self-adjoint
extension is
\begin{equation}
S_{m}^{\lambda_{m}}(k,\alpha)=
e^{2i\delta_{m}^{{\lambda_{m}}}(k,\alpha)}=
  \left[
    \frac
    {1+i\mu_{m}^{\lambda_{m}}(k,\alpha)}
    {1-i\mu_{m}^{\lambda_{m}}(k,\alpha)}
  \right]
  e^{2i\Delta_{m}(\alpha)},
\end{equation}
or by using Eq. \eqref{eq:mul}, we have
\begin{equation}
  S_{m}^{\lambda_{m}}(k,\alpha)
  =
  \left[
    \frac
    {\lambda_{m} k^{2|m-\alpha s|}
      \Gamma(1-|m-\alpha s|)e^{i\pi |m-\alpha s|}+
      4^{|m-\alpha s|}\Gamma(1+|m-\alpha s|)
    }
    {\lambda_{m} k^{2|m-\alpha s|}
      \Gamma(1-|m-\alpha s|)e^{-i\pi |m-\alpha s|}+
      4^{|m-\alpha s|}\Gamma(1+|m-\alpha s|)
    }
  \right]
  e^{2i\Delta_{m}(\alpha)}.
  \label{eq:smatrix}
\end{equation}
Hence, for any value of the self-adjoint extension parameter
$\lambda_{m}$, there is an additional scattering.
If ${\lambda_{m}}=0$, we achieve the corresponding result
for the AC problem with Dirichlet boundary condition
\cite{AP.1983.146.1}, $S_{m}^{0}(k,\alpha)=e^{2i\Delta_{m}(\alpha)}$.
For ${\lambda_{m}}=\infty$, we get
$S_{m}^{\infty}(k,\alpha)=e^{2i\Delta_{m}(\alpha)+2i\pi |m-\alpha s|}$.

In accordance with the general theory of scattering, the poles
of the $S$-matrix in the upper half of the complex plane
\cite{PRC.1999.60.34308} determine the positions of the bound
states in the energy scale.
These poles occur in the denominator of Eq. \eqref{eq:smatrix} with
the replacement $k\rightarrow i\kappa$, with $\kappa^{2}=-2ME$,
$E<0$.
Thus,
\begin{equation}
  \lambda_{m} (i\kappa)^{2|m-\alpha s|}
  \Gamma(1-|m-\alpha s|)e^{-i\pi |m-\alpha s|}+
  4^{|m-\alpha s|}\Gamma(1+|m-\alpha s|)=0.
\end{equation}
Solving the above equation for $E$, we found the bound state
energy
\begin{equation}
  E=-\frac{2}{M}
  \left[-\frac{1}{\lambda_{m}}
    \frac{\Gamma(1+|m-\alpha s|)}{\Gamma(1-|m-\alpha s|)}
  \right]^{1/|m-\alpha s|},
  \label{eq:energy_BG-sc}
\end{equation}
for $\lambda_{m}<0$.
Hence, the poles of the scattering matrix only occur for
negative values of the self-adjoint extension parameter, when we
have scattering and bound states.
In this latter case, the scattering operator can be expressed in
terms of the bound state energy
\begin{equation}
  S_{m}^{\lambda_{m}}(k,\alpha)=e^{2i\Delta_{m}(\alpha)}
  \left[
    \frac
    {e^{2 i \pi |m-\alpha s|}-(\kappa/k)^{2|m-\alpha s|}}
    {1-(\kappa/k)^{2|m-\alpha s|}}
  \right].
\end{equation}

In a previous work \cite{EPL.2013.101.51005} (cf. also
\cite{AP.2010.325.2529,PLB.2013.719.467,JMP.2012.53.122106}
for analogous systems), using another self-adjoint
extension approach, the energy bound state for the present
system was determined in terms of the physics of the problem,
and it read
\begin{equation}
  E=
  -\frac{2}{M a^{2}}
  \left[
    \left(
      \frac
      {\alpha +|m-\alpha s|}
      {\alpha -|m-\alpha s|}
    \right)
    \frac
    {\Gamma (1+|m-\alpha s|)}
    {\Gamma (1-|m-\alpha s|)}
  \right]^{{1}/{|m-\alpha s|}},
\label{eq:energy_KS}
\end{equation}
where $a$ is a very small radius smaller than the Compton
wavelength $\lambda_{C}$ of the electron
\cite{PLB.1994.333.238}, which comes from the
regularization of the $\delta$ function.
By comparing Eq. \eqref{eq:energy_BG-sc} with Eq. \eqref{eq:energy_KS}
we have
\begin{equation}
  \frac{1}{\lambda_{m}}=
  -\frac{1}{a^{2|m-\alpha s|}}
    \left(
      \frac
      {\alpha +|m-\alpha s|}
      {\alpha -|m-\alpha s|}
    \right).
\label{eq:saep}
\end{equation}
The above relation is only valid for $\lambda_{m}<0$,
consequently we have $|\alpha|\geq |m-\alpha s|$ and due to
$|m-\alpha s|<1$ it is sufficient to consider $|\alpha|\geq 1$ to
guarantee $\lambda_{m}$ to be negative.
A necessary condition for a $\delta$ function generates an
attractive potential, which is able to support bound states, is
that the coupling constant must be negative.
Thus, the existence of bound states requires
\begin{equation}
\alpha \leq -1.
\end{equation}
Also, it seems from the above equation and from Eq. \eqref{eq:deltac}
that we must have $g\upsilon\lambda < 0$ and there is a minimum value
for this product to ensure the presence of a bound state.

The scattering amplitude $f(k,\alpha)$ can be now obtained using
the standard methods of scattering theory, namely
\begin{align}
  f(k,\alpha)
  = {} &
  \frac{1}{\sqrt{2\pi i k}}\sum_{m=-\infty}^{\infty}
  \left(
    S_{m}^{\lambda_{m}}(k,\alpha)-1
  \right)
  e^{im\varphi}   \nonumber \\
  ={} &
    \frac{1}{\sqrt{2\pi i k}}
    \left\{
      \sum_{|m-\alpha s|\geq 1}
      (e^{2 i \Delta_{m}(\alpha)}-1)e^{im\varphi}
    \right. \nonumber \\
    & +
    \left.
    \sum_{|m-\alpha s|< 1}
    (e^{2 i \Delta_{m}(\alpha)}
    \left[
      \frac
      {1+ i\mu_{m}^{\lambda_{m}}(k,\alpha)}
      {1-i\mu_{m}^{\lambda_{m}}(k,\alpha)}
    \right]-1)e^{im\varphi}
  \right\}.
  \label{eq:scattamp}
\end{align}
The first sum is the AC amplitude
(i.e., when the $\delta$ function is absent),
while the second sum is the contribution that come from the singular
solutions.
In the above equation we can see that the scattering amplitude
is energy dependent (cf. Eq. \eqref{eq:mul}).
This is a clearly manifestation of the known non-conservation of
the helicity in the AC scattering \cite{PRL.1990.64.2347},
because the only length scale in the nonrelativistic problem is
set by  $1/k$, so it follows that the scattering amplitude would
be a function of the angle alone, multiplied by $1/k$
\cite{PRD.1977.16.1815}.
In fact, the failure of helicity conservation expressed in
Eq. \eqref{eq:scattamp}, it stems from the fact that the
$\delta$ function singularity make the Hamiltonian and the
helicity nonself-adjoint operators
\cite{PRD.1977.15.2287,PRL.1983.50.464,PLB.1993.298.63,
NPB.1994.419.323}.
By expressing the helicity operator,
$\hat{h}=\boldsymbol{\Sigma} \cdot \boldsymbol{\Pi}$,
in terms of the variables used in Eq. \eqref{eq:wavef}, we attain
\begin{equation}
  \hat{h} =
  \left(
    \begin{array}{cc}
      0
      & \displaystyle -i
      \left(
        \partial_{r}+\frac{|m-\alpha s|+1}{r}
      \right) \\
      \displaystyle -i
      \left(
        \partial_{r}-\frac{|m-\alpha s|}{r}
      \right)
      & 0
    \end{array}
  \right).
\end{equation}
Notice under a parity $\pi$ transformation
$\hat{h}\to\pi^{\dagger}\hat{h}\pi=-\hat{h}$, that comes
immediately from the parity transformation
$\pi^{\dagger}r\pi=-r$.
This is in fact the helicity odd-parity property.
The helicity operator share the same issue as the Hamiltonian
operator in the interval $|m+\alpha s|<1$, i.e., it is not
self-adjoint \cite{PRD.1994.49.2092,JPA.2001.34.8859}.
Despite that on a finite interval $[0,L]$, $\hat{h}$  is a
self-adjoint operator with domain in the functions satisfying
$\xi(L)=e^{i\theta}\xi{(0)}$, it does not admit a self-adjoint
extension on the interval $[0,\infty)$ \cite{AJP.2001.69.322},
and consequently it cannot be conserved, thus the helicity
conservation is broken due to the presence of the singularity at
the origin \cite{PRD.1977.16.1815,PRL.1983.50.464}.

\section{Conclusion}
\label{sec:conclusion}

We have studied the spin-1/2 AC scattering problem with a
Lorentz-violating and CPT-odd nonminimal coupling between
fermions and the gauge field in the context of the 
nonrelativistic limit of the Dirac equation.
It has been shown that there is an additional scattering for any
value of the self-adjoint extension parameter and for negative
values of this parameter there are non-zero energy bound states.
The scattering amplitude show a energy dependency, so the
helicity in not conserved.
This stem from the fact that the helicity operator is not
a self-adjoint extension operator.

\section*{Acknowledgments}
F. M. Andrade acknowleges reasearcher grants by Funda\c{c}\~{a}o
Arauc\'{a}ria project No. 205/2013 and 
E. O. Silva acknowledges researcher grants by CNPq-(Universal) project
No. 484959/2011-5.

\section*{References}
\bibliographystyle{unsrt}

\end{document}